\documentclass[%
 reprint,
 amsmath,amssymb,
 aps,
]{revtex4-2}

\usepackage{graphicx}
\usepackage{dcolumn}
\usepackage{bm}
\usepackage{physics}
\usepackage{mathrsfs}
\usepackage[cal=boondoxo]{mathalfa}

\usepackage[dvipsnames]{xcolor}

\begin{document}

\preprint{APS/123-QED}

\title{Averages in optical coherence: resolving the Magyar and Mandel-Wolf paradox}

\author{Joscelyn van der Veen}
\author{Daniel James}%
\affiliation{Department of Physics, University of Toronto, Toronto, Canada}

\date{\today}

\begin{abstract}

The ubiquity of optical coherence arising from its importance in everything from astronomy to photovoltaics means that underlying assumptions such as stationarity and ergodicity can become implicit. When these assumptions become implicit, it can appear that two different averages are independent: the finite time averaging of a detector and the ensemble average of the optical field over multiple detections. One of the two types of averaging may even be ignored. We can observe coherence as an interference fringe pattern and learn properties of the field through both methods of averaging but the coherence will not be the same, as demonstrated by the Magyar and Mandel-Wolf paradox.

\end{abstract}

\maketitle

Optical coherence is a very important concept as it arises in practically every optical imaging system. The coherence of systems has to be considered not only in applications that make use of it for measurement such as interferometry and tomography but also in any optical observations such as those in astronomy and photovoltaics where the coherence of light affects detection \cite{Korotkova2020}. Mandel and Wolf established much of optical coherence theory \cite{Mandel1995} but the topic has been further addressed by numerous well-known sources such as Goodman \cite{Goodman2015} and Gbur and Visser \cite{Gbur2022}.

There are implicit assumptions underpinning optical coherence theory. A key assumption is that the optical field is both stationary and ergodic, where stationarity refers to an invariance of the field under arbitrary time translations and ergodicity refers to the equality of information carried by every probabilistically possible realization of the field \cite{Mandel1995}. The assumption of stationarity and ergodicity allows us to separate the statistics of the field from those of the detector used to measure the field. However, this means that the separation of field and detector statistics is generally used without stating its reliance on the assumption of stationarity and ergodicity. This separation further leads to considering only the statistical average of the field or the detector without specifying that one is being ignored.

\begin{figure}
    \centering
    \includegraphics[width=\columnwidth]{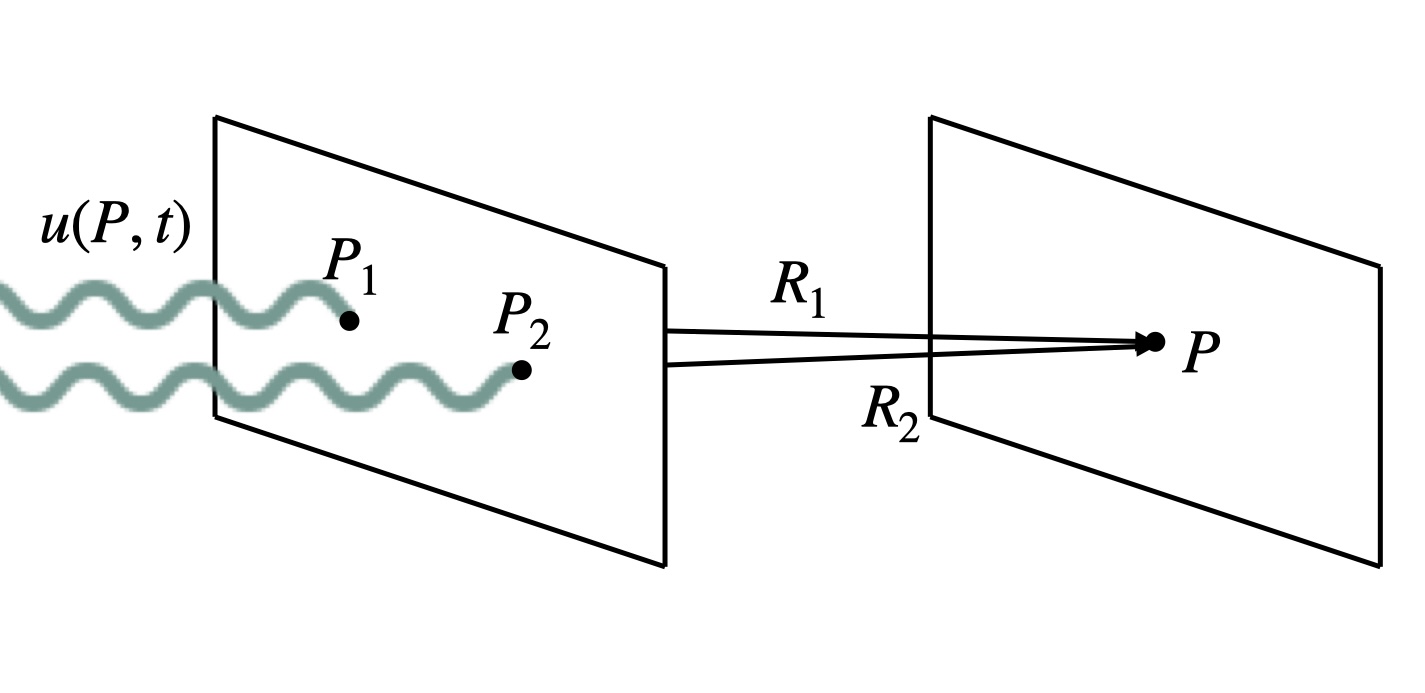}
    \caption{A Young's interference experiment that assumes the frequencies in $u(P,t)$ are short compared to the distance between the pinholes and plane of detection.}
    \label{fig:interference-diagram}
\end{figure}

The paradox between the results of Magyar and Mandel and the results of Wolf demonstrate the problem with leaving these assumptions implicit. Wolf considered a Young's interference experiment (Fig. \ref{fig:interference-diagram}) with a stationary and ergodic field being filtered narrowly in frequency at the pinholes $P_1$ and $P_2$. He derived that the visibility of the interference fringes on the screen should still be limited by the spectral coherence of the original fields at the filtered frequency \cite{Wolf1983}. Magyar and Mandel performed an experiment where they interfered two approximately monochromatic independent maser beams, which is equivalent to a Young's interference experiment filtered narrowly at the centre frequency of the maser beams. From Wolf's result, we would expect the visibility of the interfering maser beams to be very small as the independent maser beams have independent phase fluctuations and therefore low spectral coherence. However, Magyar and Mandel could measure perfect visibility with a short enough detection time \cite{MAGYAR1963}.

In this letter, we describe the difference between the ensemble average of the optical field and the finite time average of a realistic detector to show where breaking the assumption of stationarity and ergodicity would affect measurement results. We then connect the different types of averaging to the difference between amplitude and intensity coherence. Finally, we resolve the Magyar and Mandel-Wolf paradox by revealing that Magyar and Mandel defined the visibility as the intensity coherence while Wolf defined it as the amplitude coherence.

Consider a single polarization component of the electric field. We can define a complex analytic signal associated with the electric field but containing only positive frequency components. Defining the carrier frequency as $\omega_0$, the analytic signal is given by,

\begin{equation}
    u(P,t)=z(P,t)e^{-i\omega_0t}
    \label{eq:u}
\end{equation}

\noindent where $z(P,t)$ depends on the the time $t$ and the spatial point $P$ in the plane of propagation.

The signal $u(P,t)$ is quasi-monochromatic so long as the bandwidth of $z(P,t)$ is small compared to the carrier frequency $\omega_0$. This condition is generally satisfied for optical fields besides those in the few femtosecond or attosecond regime and the description of a signal in terms of a carrier frequency is only a useful description of the field in this regime \cite{Diels2006}.

If the field $u(P,t)$ originates from a realistic source such as a laser, then the envelope $z(P,t)$, and therefore $u(P,t)$, will not be predictable since there can be fluctuations in physical properties such as the phase or intensity. Realistic sources produce light that must be treated as a random process \cite{Goodman2015}.

As an example, take the envelope to be a Gaussian pulse with an amplitude $A$ that is a random variable with a probability distribution $p(A,t)$ that may change with time. Then we have $z(P,t)=Ae^{-t^2/\tau^2}$ that can be averaged in two different ways,

\begin{enumerate}
    \item The expectation value, often called the ensemble average over realizations of the field, of the envelope is 
    \begin{equation}
        \expval{z(P,t)}=\expval{A}e^{-t^2/\tau^2}
        \label{eq:time-avg-def}
    \end{equation} 
    \noindent where $\expval{A}=\int_{-\infty}^\infty Ap(A,t)\dd{A}$.
    \item The finite time average over some measurement time $T$ is 
    
    \begin{equation}
        \bar{z}(P,t)=\frac{1}{T}\int_t^{t+T}Ae^{-(t')^2/\tau^2}\dd{t'}
        \label{eq:ensemble-avg-def}
    \end{equation}
\end{enumerate}

From this example, we can see that for any envelope $z(P,t)$, taking the ensemble average over realizations of the field requires taking the expectation value of $z(P,t)$ with respect to each random variable within the function. If the probability distributions of all the random variables in $z(P,t)$ are the same at all times, which means the random variables are stationary and ergodic, then the ensemble average over realizations of the field and the finite time average over some measurement time of $z(P,t)$ are independent of one another. If any random variable is not stationary and ergodic, then taking the finite time average would change the ensemble average by considering only the probability distribution of the random variables during the measurement time. For example, if we have a collection of continuously excited dipoles, the radiation field they emit will have randomness in factors such as phase and intensity but this randomness will have the same expectation and variance regardless of the time segment we measure. However, if we had a system of something like Rydberg atoms that can blockade excitations when one becomes excited, the radiation would no longer be stationary and ergodic, and so the expectation value would depend on the time of the finite time average when both are considered.

This discussion of averages is necessary to understand what is truly measured by a detector. The output of a detector can be derived semi-classically \cite{Mandel1964} and quantum mechanically \cite{Glauber1963} as well as containing a variety of other realistic detector parameters \cite{Fleischhauer1998,Rohde2006} but they effectively agree that for the signal $u(P,t)$, a detector measures,

\begin{equation}
    \mathcal{I}(P,t) = \alpha\expval{\int_t^{t+T}u^*(P,t')u(P,t')\dd{t'}}
    \label{eq:pd-output}
\end{equation}

\noindent where $\alpha$ is the efficiency of the detector, $T$ is the integration time of the detector, and $\expval{}$ is defined in Eq. \ref{eq:ensemble-avg-def}. 

Specifically, Eq. \ref{eq:pd-output} is what a detector would measure on average after multiple instances of detection. For a single time period $T$, the detector would only measure a single value of the random variables in the field with probability given by their probability distributions and we would not have an expectation value $\expval{}$. 

To find the interference fringe visibility of a field, we would need multiple detector measurements to obtain $\mathcal{I}$ that is an average, since visibility is defined \cite[Eq. 4.3-23]{Mandel1995},

\begin{equation}
    \mathcal{V}(r)=\frac{\mathcal{I}_{max}-\mathcal{I}_{min}}{\mathcal{I}_{max}+\mathcal{I}_{min}}
    \label{eq:vis-def}
\end{equation}

The visibility is also referred to as the coherence but it is specifically an amplitude coherence that is sensitive to phase. We may instead define an intensity coherence that considers the normalized intensity difference of a single detection rather than an average of detections. Intensity coherence does not include the expectation value in $\mathcal{I}(P,t)$ and is therefore insensitive to phase fluctuations. Intensity coherence can also be used to measure properties of a signal \cite{Thekkadath2023} but it is physically and mathematically different than visibility as defined in Eq. \ref{eq:vis-def}.

Unfortunately, the same words are sometimes used to describe both intensity coherence and amplitude coherence, which can result in misleading statements or seeming contradictions. For example, in the well-known paper by Magyar and Mandel showing the interference between independent maser beams, they considered the signal time averaged over the detector measurement time $T$,

\begin{equation}
    \frac{S(t,T)}{T}=\frac{1}{T}\int_t^{t+T}\left|u\left(P_1,t-\frac{1}{2}\tau\right)+u\left(P_2,t+\frac{1}{2}\tau\right)\right|^2\dd{t'}
    \label{eq:mm-intensity}
\end{equation}

\noindent where $\tau$ is the optical path length difference between the two sources.

The time averaged signal of the interfering maser beams has a sinusoidal variation with $\tau$ producing a fringe pattern. This fringe pattern has a smaller height when the central frequency of the maser beams differ more compared to the inverse detection time. Averaging the fringe pattern from multiple time averaged signals also reduces the height difference between the maxima and minima of the fringe pattern since randomness of the maser phases mean the position of the fringe maxima and minima on the detector change with each detection.

While Magyar and Mandel acknowledge the shift in fringe position and Mandel has defined visibility in terms of the expectation value of the signal (Eq. \ref{eq:vis-def}) elsewhere \cite{Mandel1995}, in this paper they define fringe visibility by the result of a single time averaged measurement and state they can measure a perfect fringe visibility with a short enough detection time \cite{MAGYAR1963}. This result is actually stating that the intensity coherence depends only on shortness of the detection time compared to the inverse frequency difference of the sources but the amplitude coherence, which is normally what is referred to by visibility, would still be decreased by the shifting of the fringe pattern with each detection.

This lack of distinction between amplitude and intensity coherence in the Magyar and Mandel paper can lead to misleading descriptions of the results. A recent review of the experiment states, ``According to the classical wave theory, [the two maser beams] will not produce a pattern on average because of random fluctuations between them; however, over a short time period, the two fields will be in phase and produce interference.'' \cite{Gbur2022}. This conflates the two requirements for perfect interference: that the measurement time must be short compared to the inverse frequency difference between the two maser beams and that the interference must be a single measurement rather than an average of measurements. 

The use of visibility to refer to intensity coherence also leads to the Magyar and Mandel-Wolf paradox. Wolf used amplitude coherence as the definition for visibility when he derived that the coherence of narrowband interference is limited by the spectral coherence $\mu$ at the filter wavelength $\omega_0$ \cite{Wolf1983},

\begin{equation}
    \mu(P_1,P_2,\omega_0)=\frac{W(P_1,P_2,\omega_0)}{\sqrt{W(P_1,P_1,\omega_0)W(P_2,P_2,\omega_0)}}
    \label{eq:wolf-spec-coh}
\end{equation}

\noindent where $W(P_1,P_2,\omega_0)$ is a cross-spectral density that is an expectation value $\expval{}$ of the field.

Since the cross-spectral density is an expectation value over multiple measurements, the spectral coherence of the Magyar and Mandel experiment would be small due to the randomness of the phases. Thus Wolf's result predicts low amplitude coherence for the Magyar and Mandel experiment, which agrees with their measured result of a randomly shifting fringe pattern that reduces fringe visibility on average.

We can also confirm that the setup for Wolf's derivation agrees with Magyar and Mandel's experimental result for intensity coherence. We can examine this in Wolf's setup by considering a finite detection time.

Filtering the field given in Eq. \ref{eq:u} narrowly in frequency we have, 

\begin{equation}
    u_f(P,t)=\int_{-\infty}^{\infty}Z(P,\omega-\omega_0)F(\omega-\omega_f)e^{-i\omega t}\dd{\omega}
\end{equation}

\noindent where $F(\omega-\omega_f)$ is the filter spectrum centred at $\omega_f$ and $Z(P,\omega)$ is the Fourier transform of the envelope $z(P,t)$.

So long as the bandwidth of the filter is narrow compared to the bandwidth of the slowly varying envelope $z(P,t)$, 

\begin{equation}
    u_f(P,t)=Z(P,\omega_f-\omega_0)f(t)e^{-i\omega_ft}
\end{equation}

\noindent where $f(t)$ is the inverse Fourier transform of the filter function.

When the filters are very narrowband, small differences in the central frequency of the two filters become non-negligible. We can therefore find the amplitude coherence (Eq. \ref{eq:vis-def}) for the Young's interference experiment in Fig. \ref{fig:interference-diagram} with the field filtered at the pinholes $P_1$ and $P_2$ at frequencies $\omega_f$ and $\omega_f+\delta\omega$ respectively. 

\begin{equation}
    \mathcal{V}(P)=\left|\mu(P_1,P_2,\omega_f-\omega_0,\delta\omega)\right|\left|\chi(\tilde{t},T,\delta\omega)\right|
    \label{eq:filtered-vis}
\end{equation}

\noindent where $\mathcal{V}(P)$ is a product of the spectral coherence

\begin{multline}
    \mu(P_1,P_2,\omega_f-\omega_0,\omega) \\
    =\frac{\expval{Z^*(P_2,\omega_f-\omega_0)Z(P_1,\omega_f+\delta\omega-\omega_0)}}{\sqrt{\expval{|Z(P_1,\omega_f+\delta\omega-\omega_0)|^2}\expval{|Z(P_2,\omega_f-\omega_0)|^2}}}
    \label{eq:spectral-coherence}
\end{multline}

\noindent and a part dependent on the filter function

\begin{equation}
    \chi(\tilde{t},T,\Delta\omega_f)=\frac{\int_{\tilde{t}}^{\tilde{t}+T}|f(t')|^2e^{-i\delta\omega t'}\dd{t'}}{\int_{\tilde{t}}^{\tilde{t}+T}|f(t')|^2\dd{t'}} 
    \label{eq:xi}
\end{equation}

This assumes we have a stationary and ergodic field such that the expectation value $\expval{}$ and the finite time average are independent.

\begin{figure}
    \centering
    \includegraphics[width=\columnwidth]{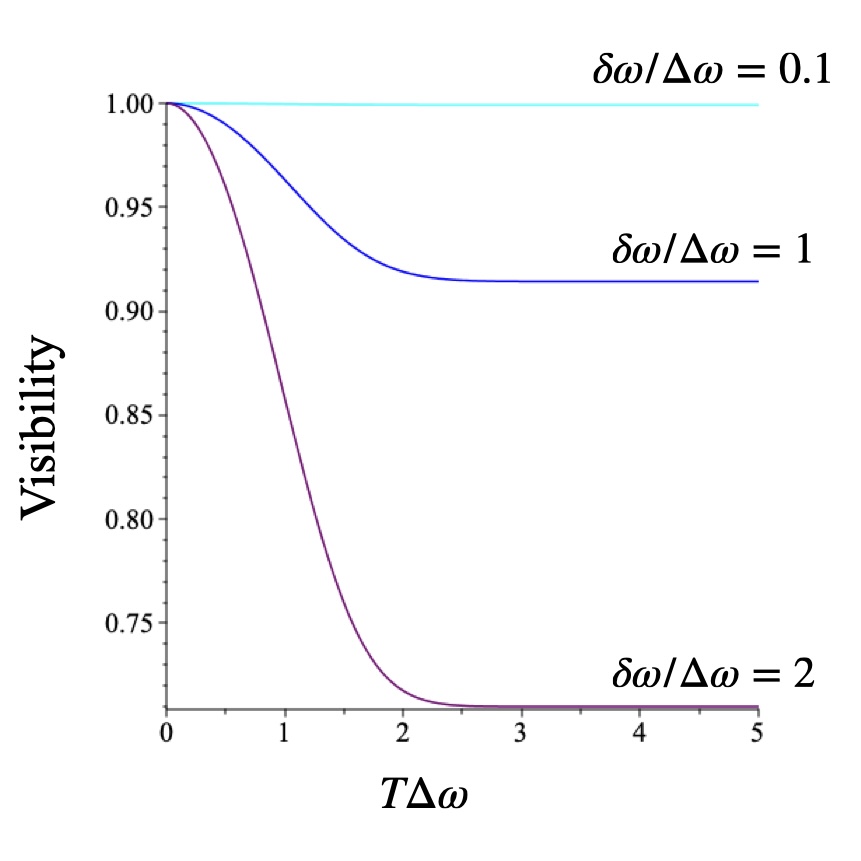}
    \caption{Visibility of interference fringes from a spatially coherent source that has been filtered narrowly in frequency at the pinholes.}
    \label{fig:filtered-vis}
\end{figure}

Eq. \ref{eq:filtered-vis} shows that the visibility of narrowband filtered interference is limited by the degree of spectral coherence $\mu$ as shown by Wolf. By including the finite measurement time of the detector, we can also see the visibility depends on the relationship between the filter frequency difference $\delta\omega$ and the measurement time regardless of expectation value, as Magyar and Mandel showed in their intensity interference experiment. As an example, we can consider a Gaussian filter function with bandwidth $\Delta\omega$ and see from Fig. \ref{fig:filtered-vis} how small differences in the filters' central frequency can reduce visibility when the bandwidth is very narrow, but this can also be compensated by measuring for shorter detection times. 

Since accurate measurements of spectral coherence are of interest in current applications such as photovoltaics \cite{Divitt2015}, it's useful to have not only a limitation on the narrowness of filtering but also a way to compensate for narrow filtering even with an uncertainty in the filter frequency by measuring for shorter times.

In summary, when measuring a field, we can take multiple measurements to observe the statistical properties of the field as an expectation value but each measurement also has some finite time averaging over the duration of the detector's measurement. We can consider these averages independently when the field is stationary and ergodic. Visibility is usually defined as the amplitude coherence, which has an expectation value so it is an average over many measurements. However, the intensity coherence, which is a single measurement, may also be referred to as visibility. 

Magyar and Mandel referred to an intensity coherence as the visibility when they measured the interference between two independent maser beams. This terminology leads to descriptions of the experiment that can imply they measured perfect amplitude coherence and to contradictions with other results, such as the Magyar and Mandel-Wolf paradox. By leaving the assumptions of optical coherence implicit, we can falsely equate different types of coherence. We should not only ensure that future results in optical coherence specify the quantities over which they average but also carefully review the type of coherence actually used in any previous results.

We wish to acknowledge Joseph Eberly and Rayf Shiel for useful conversation that inspired this paper as well as Zeke Horsley and Zhuoran Bao for useful correspondence and discussion. We also acknowledge the support of the Natural Sciences and Engineering Research Council of Canada (NSERC).

\bibliography{references}

\end{document}